\documentclass[aps,prl,twocolumn,superscriptaddress,showpacs]{revtex4}
\usepackage{graphicx}  
\usepackage{dcolumn}   
\usepackage{bm}        
\usepackage{amssymb}   
\usepackage{amsmath}
\usepackage{verbatim}
\usepackage{color}

\begin{document}

\title{Demonstration of zero optical backscattering from single
  nanoparticles}

\author{Steven Person}
\affiliation{Institute of Optics, University of Rochester,
Rochester, New York 14627, USA}  
\author{Manish Jain}
\affiliation{Institute of Optics, University of Rochester,
Rochester, New York 14627, USA}
\author{Zachary Lapin}
\affiliation{Institute of Optics, University of Rochester,
Rochester, New York 14627, USA}
\affiliation{ETH Z{\"u}rich, Photonics Laboratory, 8093 Z{\"u}rich, Switzerland}
\author{Juan Jose S{\' a}enz}
\affiliation{Departamento de F\'isica de la Materia Condensada, Instituto ``Nicol\'as Cabrera'' and  Centro de Investigaci\'on en F\'isica de la Materia Condensada (IFIMAC),
Universidad Aut\'onoma de Madrid, 28049-Madrid, Spain.}
\author{Gary Wicks}
\affiliation{Institute of Optics, University of Rochester,
Rochester, New York 14627, USA}
\author{Lukas Novotny}
\affiliation{Institute of Optics, University of Rochester,
Rochester, New York 14627, USA}
\affiliation{ETH Z{\"u}rich, Photonics Laboratory, 8093 Z{\"u}rich, Switzerland}
\date{\today}

\date{\today}

\begin{abstract}
    We present the first experimental demonstration of zero backscattering from
    nanoparticles at {\em optical} frequencies as originally discussed by Kerker
    {\it et. al.} [M. Kerker, D. Wang, and C. Giles, J. Opt. Soc. A {\bf 73}, 765
    (1983)]. GaAs pillars were fabricated on a fused silica substrate and
    the spectrum of the backscattered radiation was measured in the wavelength range 600-1000 nm. Suppression of
     backscattering occurred at $\sim$725 nm, agreeing with calculations based on the discrete
    dipole approximation. Particles with zero backscattering provide new functionality for metamaterials
and optical antennas.
\end{abstract}

\maketitle

Nanofabrication techniques continue to fuel the search for
structures that exhibit unique optical properties. Among these structures 
are optical antennas, which concentrate and
direct light~\cite{bharadwaj09a}, and metamaterials that can
mimic optical constants not found in
nature~\cite{shalaev04,soukoulis11}. Optical antennas and metamaterials often
capitalize on the plasmonic properties of metal nanostructures~\cite{schuller10,kawata09}. Plasmons,
however, have large non-radiative losses that restrict practical
applications to narrow frequency bands~\cite{stockman07b}, even if active gain media are employed~\cite{Hess12}. Recently, alternative approaches using dielectric
materials have been proposed for constructing optical
antennas~\cite{schuller09b,rolly12a,liu12,devilez10} and
metamaterials~\cite{schuller07,zhao09,jylha06}.\\[-2ex]

The optical properties of dielectric nanostructures are strongly influenced by the interaction of electric
and magnetic Mie resonances. Thirty years ago, Kerker {\it et. al.} 
showed that the
backscattered light from spherical scatterers can be completely suppressed if the 
dielectric  and magnetic properties of the scatterers are the same ($\varepsilon = \mu$)~\cite{kerker83}. A particle with these properties
exhibits equal electric ($a_{\rm n}$) and magnetic ($b_{\rm n}$) multipole
coefficients~\cite{bohren83} that destructively interfere in the backward
propagating direction (first Kerker condition). Until recently, zero backscattering at
visible wavelengths has been a theoretical
curiosity~\cite{garcia-camara08,garcia-camara11,garcia-camara10} due to
the lack of magnetic materials ($\mu \neq 1$) in the optical regime.\\[-2ex]

Kerker's condition for zero backscattering can be generalized for
spherical particles assuming that the scattered field is sufficiently described by the
electric ($a_1$) and magnetic ($b_1$) dipole terms of the Mie expansion
\cite{nieto-vesperinas11}. In terms of $a_1$ and $b_1$ the electric and
magnetic polarizabilities are defined as
\begin{equation}
    \alpha_{\rm e} = \frac{3 i \varepsilon}{2k^3}a_1  \qquad
    \alpha_{\rm m} = \frac{3 i}{2 \mu\:\! k^3}b_1
\end{equation} 
respectively, where $k=2\pi n_{\rm m}/ \lambda$, with $n_{\rm m}$ being the
refractive index of the surrounding medium and $\lambda$ the vacuum wavelength. The
two dipoles interfere constructively or destructively depending on their
relative phase. In the backward direction the particle's scattering cross-section
is
\begin{equation}
    \label{alpha_eq}
    \sigma_{\mathrm{b}} \; = \;
    4 \pi k^4 I_{\alpha}\Big[1 + V_{\alpha}\cos(\pi - \Delta\phi_{\alpha} )\Big]
\end{equation}  
where $I_{\alpha}\; = \; |\alpha_{\rm e} / \varepsilon|^2 +
|\mu\;\!\alpha_{\rm m}|^2$ is the incoherent sum of the two dipoles,
$\Delta \phi_{\alpha}$ is the phase difference between the dipoles, and
$V_{\alpha}$, given by
\begin{equation}
    V_{\alpha} \; = \; \frac{2\;\!|\alpha_{\rm e}/\varepsilon|\;\!
      |\mu\;\!\alpha_{\rm m}|}{I_{\alpha}}, 
\end{equation}  
controls the backscattering suppression. A minimum in the backscattered
field will occur whenever the electric and magnetic dipoles are oscillating
in phase ($\Delta \phi_{\alpha}=0$). The minimum will be exactly zero if
$V_{\alpha}$ is equal to one. \\[-2ex]

\begin{figure}[!b]
\includegraphics[width=25em]{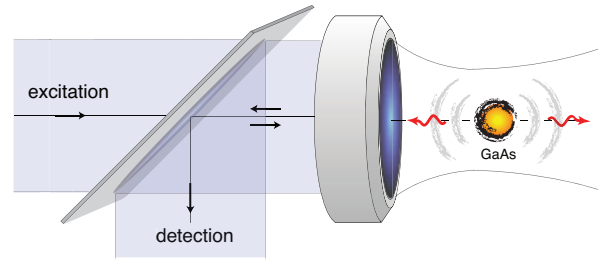}
    \caption{Illustration of the backscattering measurement. White light is weakly focused on a GaAs particle of radius $\sim$ 90 nm. Backscattered light is separated by a 50/50 beamsplitter and sent to a spectrometer.
    \label{figure1}}
\end{figure}
Note that the elimination of the backscattered field in Eq. (\ref{alpha_eq}) is 
valid even for non-magnetic materials ($\mu=1$) provided that the dipole approximation
holds. This approximation requires dielectric nanoparticles with a large index of refraction ($n \sim
3-4$), such as silicon and germanium~\cite{gomez-medina11,garcia-etxarri11}. 
Recently, the electric and magnetic dipole and
quadrupole resonances have been experimentally measured for silicon
spheres in the visible~\cite{kuznetsov12,evlyukhin12,wheeler09}. The
large refractive index of these nanoparticles make them ideally suited for verifying Kerker's  
theoretical predictions.\\[-2ex]

\begin{figure}[!b]
    \centering\includegraphics[width=25em]{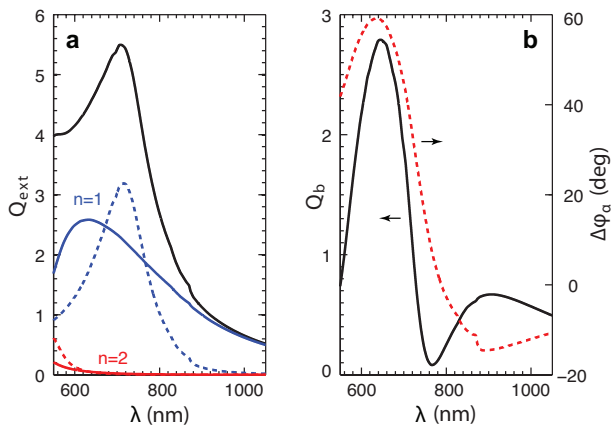}
    \caption{Mie theory calculations of a 90 nm radius sphere of GaAs
      embedded in a uniform medium of refractive index $n_{\rm
        m}=1.47$. (a) Electric (solid) and magnetic (dashed) dipole (n=1)
      and quadrapole (n=2) contributions to the total extinction efficiency
      ($Q_{\rm ext}$). (b) Backscattering efficiency ($Q_{\rm
        b}$) and phase difference ($\Delta \phi_{\alpha}$) between the
      electric  and magnetic dipoles.
    \label{figure2}}
\end{figure}

In this letter we experimentally demonstrate the 
suppression of optical backscattering from dielectric nanoparticles, as
predicted by Kerker~\cite{kerker83}. A recent experiment has demonstrated suppressed backscattering at microwave frequencies~\cite{geffrin12},
however, a verification at optical wavelengths has not yet been performed. We chose
GaAs as a scattering material due to its large index of refraction
($n\sim\,3.6$) in the 600 to 1000 nm wavelength
range~\cite{palik98}. The experimental arrangement is sketched in Fig.~\ref{figure1}: white light is weakly focused on a sample of well separated GaAs nanoparticles and the backscattered light is separated by a beamsplitter and directed on a spectrometer.  Fig.~\ref{figure2}(a) is a plot of the extinction
efficiency ($Q_{\rm ext} = \sigma_{\rm ext}/ \pi r^2$) for a GaAs sphere of radius $r$=90 nm  immersed in a medium of index $n_{\rm m} = 1.47$ . The
contributions of the dipole terms ($a_1$, $b_1$) and the quadrupole terms ($a_2$, $b_2$)
of $Q_{\rm ext}$ are  shown as separate curves. The curves show that for wavelengths longer than $\sim$600 nm
 the electric and magnetic quadrupole terms can be neglected. The backscattering efficiency ($Q_{\rm b}=\sigma_{\rm b}/\pi
r^2$) is plotted in Fig.~\ref{figure2}(b). A minimum in $Q_{\rm b}$ is
located at $\sim$775 nm where the relative phase between the electric and
magnetic dipoles crosses zero $(\Delta\phi_{\alpha}\sim 0)$.\\[-2ex]

\begin{figure}[!t]
\centering\includegraphics[width=24em]{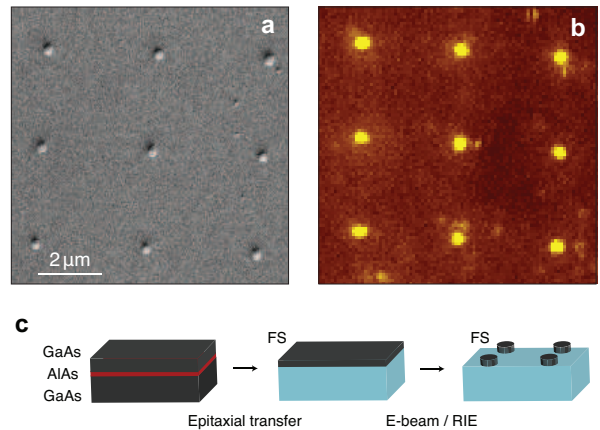}
    \caption{(a,b) Characterization of the fabricated GaAs nanoparticle sample.
    (a) SEM image of an array of GaAs pillars of radius $r\sim$ 90 nm. (b) Confocal photoluminescence image of the
      same sample area.  Wavelength of excitation laser: $\lambda=532$ nm.
   (c) Steps of the sample fabrication process showing the
      transfer of the GaAs epitaxial layer to a fused silica (FS) substrate
      and the generation of GaAs pillars with {\it e}-beam lithography and
      reactive-ion etching (RIE).
    \label{figure3}}
\end{figure}

To measure backscattering from GaAs nanoparticles embedded in a uniform 
environment we implemented an epitaxial
lift-off technique in conjunction with a water-bonding procedure to attach
a high quality GaAs membrane (grown on a GaAs substrate) to a fused silica
substrate~\cite{yablonovitch87,demeester93}. Direct growth of 
GaAs  on fused silica is avoided because it results in a high density 
of dislocations. Figure~\ref{figure3}(b) illustrates the fabrication steps. Using molecular beam
epitaxy (MBE), a 40 nm sacrificial layer of AlAs was deposited on a GaAs
substrate. On top of the AlAs layer a 1 $\mu$m film of GaAs was
grown. Using the epitaxial lift-off procedure, the 1 $\mu m$ GaAs film was
transfered to a fused silica substrate. The transfered GaAs film was
initially reduced to a thickness of $\sim$150 nm by reactive ion etching
(RIE). An array of $\sim$175 nm diameter discs was then patterned in a
Poly(methyl methacrylate) (PMMA) resist using {\it e}-beam
lithography. After developing the resist, GaAs pillars were created by
RIE. The leftover PMMA was removed with an additional oxygen plasma
etch. Fig.~\ref{figure3}(a) shows a scanning electron microscope (SEM)
image of the fabricated structures. To verify that no residual GaAs
remained after etching, a confocal scanning image of the photoluminescence
was taken with an excitation laser of $\lambda=532$ nm. The confocal  image
in Fig.~\ref{figure3}(b) shows strong signal from the GaAs pillars with
limited luminescence from the fused silica substrate.\\[-2ex]

Backscattering measurements  were done using a standard inverted microscope with
a fiber-coupled tungsten halogen white light source (c.f Fig.~\ref{figure1}). To create a uniform
environment, and eliminate any back-reflections, index matching oil covered
the top of the sample. The sample was broadly illuminated by weakly
focusing the white light source through an oil immersion objective. The
backscattered light was collected with the same objective and sent to
either an avalanche photodiode (APD) or a spectrometer. A 10 nm bandpass
filter was placed in front of the APD and the sample was scanned by a
piezo positioning stage to form an image. Spectra of single GaAs
pillars were acquired in the wavelength range of 600-1000 nm. The
collection volume was limited either by the APD detector area or the spectrometer
silt width. This ensured that light from only a single scatterer was detected even though the
sample was broadly illuminated.\\[-2ex]

The recorded spectra and APD images were corrected for source spectrum
non-uniformity, transmission losses, and detector efficiencies.  A baseline spectrum [$I_{\rm mirror}(\lambda)$] was
recorded by replacing the sample with a mirror. Additionally, the background
spectrum [$I_{\rm bkg}(\lambda)$] measured on the fused silica substrate away from the GaAs pillars
was subtracted. This background resulted from a slight refractive index
mismatch between substrate and index-matching oil. The final
backscattering efficiency is calculated as
\begin{equation}
    Q_{\rm b}(\lambda) \; = \; \beta\, \frac{I_{\rm meas}(\lambda) - I_{\rm bkg}(\lambda)}{I_{\rm mirror}(\lambda)}, 
\end{equation}  
where $I_{\rm meas}$ is the recorded spectrum. The pre-factor $\beta$ is a calibration scaling factor.\\[-2ex] 

\begin{figure}[!b]
\centering\includegraphics[width=25em]{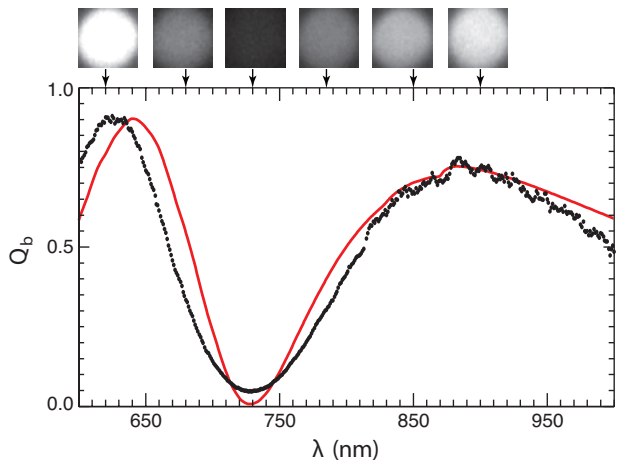}
    \caption{Backscattering spectrum from a single GaAs pillar. The black dotted curve is
      the measured spectrum and the red curve is
      the theoretical spectrum calculated using the discrete dipole approximation. The series of images above the graph are
      backscattering images of a GaAs pillar
      recorded with 10 nm bandpass filters at various center wavelengths.
    \label{figure4}}
\end{figure}

Figure \ref{figure4} shows the resulting backscattering efficiency  recorded for a
single GaAs pillar. The spectrum has a pronounced dip at
$\sim$725 nm where the backscattered field is suppressed due to the
interference between the electric and magnetic dipole components. Though
the suppressed backscattered field is obvious, there are two notable
differences in the spectral shape when comparing to the Mie theory of
Fig. \ref{figure2}. First, the scattering amplitude below $\sim$700 nm is
suppressed and, second, the location of the minimum is blue shifted. Both
of the differences are attributed to the fact that the structure is a
pillar rather than a sphere. Compared to a sphere of the same volume, a
pillar with an aspect ratio (height/diameter) below one will have the
location of the minimum blue shifted and the scattering peak to the left of
the minimum suppressed (See Supplementary Material \cite{suppl}). The
solid curve in Fig. \ref{figure4} is a simulation of the backscattered
field of a  GaAs pillar calculated using the discrete dipole approximation
\cite{draine94}. The simulated structure has a height of 137 nm and a
diameter of 165 nm, which is within 10\% of the initial design
parameters. The sequence of images above Fig.~\ref{figure4} are backscattering images with bandwidth of 10 nm and center wavelengths 
indicated by the arrows. At the
location of the spectral minimum the image of the scatterer fades into the
background.\\[-2ex]

In conclusion, we have shown the first experimental verification of
Kerker's theoretical prediction of zero backscattering in the {\em optical}
wavelength range. Using pillars, rather than spheres, slight shifts of the
wavelength of minimum backscattering can be observed. This shift as well as the reshaping of the spectrum 
are reproduced with calculations based on the discrete dipole approximation. Understanding
and verifying the response of single dielectric nanoparticles is crucial in
any bottom-up nanofabrication technique. The remarkable properties of these particles, with  zero backscattering but significant light dispersion in forward direction, suggest intriguing technological applications, for example as light diffusing elements in solar cells. Using similar GaAs structures,
experiments involving arrays of scatterers~\cite{evlyukhin10} or dielectric
antennas manipulating magnetic dipoles~\cite{schmidt12,rolly12b} are
possible. \\

\begin{acknowledgments}
This work was funded by the U.S. Department of
Energy (Grant No. DE-FG02-01ER15204). 
\end{acknowledgments}


\end{document}